\documentclass[11pt,a4paper]{article}

\usepackage{amsmath,amsthm,amssymb,amsfonts}
\usepackage{mathrsfs}
\usepackage{authblk}
\usepackage{setspace}
\usepackage{mathrsfs}
\usepackage{cases}
\usepackage{graphicx}
\usepackage{amsthm}
\usepackage[usenames]{xcolor}
\usepackage{hyperref}
\usepackage{doi}

\numberwithin{equation}{section}
\newcommand{\sfx}{{\sf{x}}}

\DeclareMathOperator{\sgn}{sgn}

\DeclareMathOperator{\arsinh}{arsinh}

\begin{document}

\title{Thermality from a Rindler quench}
\author{Jorma Louko}

\affil{School of Mathematical Sciences, University of Nottingham,\\
Nottingham NG7 2RD, UK\\
jorma.louko@nottingham.ac.uk}

\date{July 2018; revised September 2018\footnote{This 
is a peer-reviewed, un-copyedited version of an article published in Class.\ Quant.\ Grav.\  {\bf 35}, 205006 (2018). 
IOP Publishing Ltd is not responsible for any errors or omissions in this version 
of the manuscript or any version derived from it. 
The Version of Record is available online at doi:10.1088/1361-6382/aadb34. 
--- Update November 2026: typo $\eta\to t$ in \eqref{eq:KGmodes-before} corrected.}}

\maketitle

\begin{abstract}
Ultracold fermionic atoms in an optical lattice, 
with a sudden position-dependent change (a~quench) 
in the effective dispersion relation, have been 
proposed by Rodr\'iguez-Laguna \textit{et al.\ }as 
an analogue spacetime test of the Unruh effect. 
We provide new support for this analogue 
by analysing a massless scalar field on
a $(1+1)$-dimensional continuum spacetime with 
a similar quench: 
an early time Minkowski region is joined at a constant time surface, 
representing the quench, 
to a late time static region in which 
left and right asymptotically Rindler domains are 
connected by a smooth negative curvature bridge. 
We show that the quench is energetically mild, 
and late time static observers, modelled as a 
derivative-coupling Unruh-DeWitt detector, 
see thermality, 
in a temperature that equals the Unruh temperature for observers 
in the asymptotic Rindler domains. 
The Unruh effect hence prevails,  
despite the energy injected into the field by the quench and despite 
the absence of a late time Killing horizon. 
These results strengthen the motivation to realise the experimental proposal.
\end{abstract}

\singlespacing

\section{Introduction\label{sec:intro}} 

In relativistic quantum field theory, an 
observer's measurements of a quantum field 
depend on the observer's motion. 
A~celebrated example is the 
Unruh effect~\cite{Fulling:1972md,Davies:1974th,Unruh:1976db}, 
in which a linearly uniformly accelerated observer in Minkowski spacetime 
reacts to a field in its Minkowski vacuum 
by excitations and de-excitations characteristic 
of a thermal state, in the Unruh temperature~$a\hbar/(2\pi c k_B)$, 
where $a$ is the observer's proper acceleration 
(for textbooks and reviews, see~\cite{Birrell:1982ix,Crispino:2007eb,Fulling:2014wzx}). 
An experimental confirmation of the Unruh effect has remained elusive, 
due to the required magnitude of acceleration (for a discussion of the magnitudes, 
and a proposal to enhance the effect through the Berry phase, see~\cite{MartinMartinez:2010sg}). 
A~related effect exists for nonlinear uniform accelerations~\cite{Letaw:1980yv}, 
including circular motion~\cite{Letaw:1979wy,Doukas:2010wt,Jin:2014coa,Jin:2014spa}, 
and the circular motion version is related to 
the spin depolarisation of particle beams in accelerator storage 
rings~\cite{Bell:1982qr,Bell:1986ir,Leinaas:1998tu,Unruh:1998gq}, 
originally predicted by different methods \cite{Sokolov:1963zn,Jackson:1975qi} 
and observed~\cite{Johnson:1982rb}, 
but also here establishing a direct connection between 
the observation and the circular motion 
Unruh effect has remained elusive~\cite{Unruh:1998gq}. 
The prospects to observe versions of the Unruh effect with high-power 
laser systems are discussed in~\cite{Chen:1998kp,Mourou:2006zz,Brodin:2007vf,gregori-scirep}. 
An experimental confirmation of the Unruh effect would be significant 
since the mathematics underpinning the effect is 
closely related to the mathematics in 
Hawking's prediction of black hole radiation \cite{Hawking:1974sw} 
and to the mathematics of the early universe quantum effects 
that may be responsible for the origin 
of structure in the present-day Universe~\cite{Parker:1969au,mukha-wini-book}. 

It is by now well recognised that classical and quantum field theory phenomena 
in relativistic spacetimes can be simulated in laboratory systems 
described by mathematically similar 
effective field theories, classical or quantum~\cite{Barcelo:2005fc}. 
Recent experimental work includes the observation of a classical 
mode conversion that underlies 
the Hawking effect in the quantum theory~\cite{Weinfurtner:2010nu}, 
the observation of classical superradiance~\cite{Torres:2016iee}, 
the observation of quantum phenomena characteristic 
of an expanding cosmology~\cite{Eckel:2017uqx}, 
and observations interpreted as analogue 
Hawking radiation~\cite{Belgiorno:2010wn,Steinhauer:2015saa}. 
Laboratory analogues of the Unruh effect have been proposed 
in a Bose-Einstein condensate \cite{Retzker-circular} 
and in ultracold fermionic atoms 
in an optical lattice~\cite{Rodriguez-Laguna:2016kri,Kosior:2018vgx}, 
and related proposals are discussed in~\cite{Cozzella:2017ckb,Leonhardt:2017lwm}. 
A~laboratory analogue of the Gibbons-Hawking effect, 
a curved spacetime counterpart of the Unruh effect, 
has been proposed in~\cite{Fedichev:2003id,Fedichev:2003dj}. 

The purpose of this paper is to provide new evidence that the 
optical lattice proposal of \cite{Rodriguez-Laguna:2016kri,Kosior:2018vgx} 
has the requisite properties to simulate the Unruh effect, 
despite having energetic and causal properties that differ 
from those in the usual setting of the Unruh effect. 

The system analysed in \cite{Rodriguez-Laguna:2016kri,Kosior:2018vgx} 
consists of fermionic atoms held in an optical lattice, 
with a dispersion relation that can be adjusted 
to depend on both space and time. 
The detailed experimental implementation is described in~\cite{Rodriguez-Laguna:2016kri,Kosior:2018vgx}. 
Mathematically, the system is a spatially discretised fermionic 
field in an effective $(2+1)$-dimensional spacetime whose 
spatial sections are flat but the time-time component of the metric, 
determining the effective dispersion relation, may depend on both space and time. 
To simulate the Unruh effect, the spacetime metric is engineered to undergo a sudden change, 
a quench, 
from the $(2+1)$-dimensional Minkowski metric to a metric given 
by the $(1+1)$-dimensional part 
\begin{align}
ds^2 = - \chi^2 d\eta^2 + d\chi^2 
\label{eq:unregrindler}
\end{align}
plus one flat spatial dimension. 
(We set from now on $c = \hbar = k_B=1$.) 
It is shown in \cite{Rodriguez-Laguna:2016kri,Kosior:2018vgx} 
by a combination of analytic and numerical methods
that the field's behaviour at constant~$\chi$, 
sufficiently far from $\chi=0$ in terms of the lattice scale, 
has thermal characteristics, in a position-dependent 
temperature that approximates $1/(2\pi|\chi|)$. 
This thermality is interpreted as an analogue of the Unruh effect, 
on the grounds that the regions $\chi>0$ and $\chi<0$ of \eqref{eq:unregrindler}
each cover one Rindler wedge of Minkowski spacetime, 
the worldlines of constant $\chi\ne0$ have proper acceleration~$1/|\chi|$, 
and the usual Unruh effect states that an observer in a 
Rindler wedge at constant $\chi$ experiences the Minkowski 
vacuum as thermal at the Unruh temperature 
$1/(2\pi|\chi|)$~\cite{Fulling:1972md,Davies:1974th,Unruh:1976db,Birrell:1982ix,Crispino:2007eb,Fulling:2014wzx}. 

The optical lattice quench and the usual Unruh effect setup 
have however three qualitative differences, 
each of which could potentially limit 
the ability of the lattice to simulate the Unruh effect. 
First, in the usual setup the field is prepared in 
its Minkowski vacuum, implying in particular that 
the field's stress-energy tensor has a vanishing 
expectation value 
(see \cite{Buchholz:2014jta,Buchholz:2015fqa} 
for a recent reinforcement of this point). 
By contrast, a sudden quench can be expected to inject 
energy into the field, potentially lots of it, 
and the spatial inhomogeneity of the quench 
suggests that the post-quench energy may not remain static:  
does the post-quench stress-energy tensor remain small, 
in some controllable sense? 
Second, the degeneracy of \eqref{eq:unregrindler} at $\chi=0$ 
gets regularised in 
\cite{Rodriguez-Laguna:2016kri,Kosior:2018vgx} 
in terms of the spatial lattice. 
How sensitive are the Unruh effect results to this regularisation, which 
does not feature in the usual setup? 
Third, the lattice-regularised metric does not have a counterpart 
of the Minkowski spacetime future and past quadrants that
join the two Rindler wedges in the usual setup. 
Yet the arrangement of the four quadrants
is essential when the thermality in the Unruh effect is described 
in terms of the entanglement between two opposing Rindler wedges 
\cite{Fulling:1972md,Davies:1974th,Unruh:1976db,Birrell:1982ix,Crispino:2007eb,Fulling:2014wzx}, 
and new phenomena emerge when this arrangement is modified~\cite{Salton:2014jaa}. 
Does the quench still create spacelike entanglement, similar to that in the usual setup? 

In this paper we address the first two of these questions in a simplified, 
analytically solvable quench model that shares the 
potentially troublesome features of the optical lattice. 
We consider a massless scalar field in the $(1+1)$-dimensional continuum spacetime 
in which the singularity of \eqref{eq:unregrindler} at $\chi=0$ is regularised by 
modifying the $\eta\eta$ component to remain negative everywhere, 
including at $\chi=0$. 
We find that the post-quench renormalised stress-energy tensor is well 
defined and nonvanishing, 
and the `total energy', defined as the integral 
of the energy density over the constant $\eta$ surface, 
is finite. 
When the regulator is small, in a sense that we describe, 
the stress-energy tensor 
is small everywhere except in a narrow region near $\chi=0$. 
In this sense, the regularised quench is energetically mild. 
We then probe the thermality of the post-quench region by an 
Unruh-DeWitt detector on a worldline of constant~$\chi$. 
The detector's late time response is 
Planckian, in a $\chi$-dependent temperature, 
and for nonzero $\chi$ this temperature approaches 
the Unruh temperature when the regulator is taken to zero; 
further, when the regulator is small, 
the response is Planckian in the usual Unruh temperature 
also at times shortly after the quench. 
In this sense, the Unruh-type thermality in the post-quench 
region is relatively insensitive to the regulator. 

In short, we find that the Unruh effect prevails, 
despite the energy injected into the field by the quench 
and despite the absence of a late time Killing horizon. 
While our model is simplified, 
these results do strengthen the motivation to 
realise the experimental proposal of~\cite{Rodriguez-Laguna:2016kri,Kosior:2018vgx}. 

Given our results, particularly our closed expression 
for the post-quench Wightman function, 
post-quench spacelike entanglement could be investigated 
by the harvesting techniques of 
\cite{Salton:2014jaa,valentini1991,Reznik:2002fz,Reznik:2003mnx,Pozas-Kerstjens:2015gta}. 
We shall comment on the prospects and challenges of such harvesting in Section~\ref{sec:discussion}. 

The plan of the paper is as follows: 
We begin by introducing in Section \ref{sec:regdoublerindler}
a regularisation of the double Rindler metric~\eqref{eq:unregrindler}. 
Section \ref{sec:quench}
presents the regularised quench,
and quantises the scalar field individually in the pre-quench and post-quench regions. 
The Wightman function in the post-quench region, with the field prepared 
in the Minkowski vacuum of the pre-quench region, 
is evaluated in Section~\ref{sec:wightman}. 
The stress-energy tensor is analysed in Section \ref{sec:stress-energy}
and the response of an Unruh-DeWitt detector in Section~\ref{sec:thermality}. 
Section \ref{sec:discussion} presents a summary and a brief discussion. 

Spacetime points are denoted by sans serif letters. 
Overline denotes complex conjugate and dagger Hermitian conjugate. 
$\sgn(x)$ denotes the signum function, 
equal to $1$ for $x>0$, $-1$ for $x<0$, and $0$ for $x=0$.

\section{Regularised double Rindler\label{sec:regdoublerindler}}

We consider the spacetime 
\begin{align}
ds^2 = - \bigl(b^2 + \chi^2\bigr) d\eta^2 + d\chi^2
\ , 
\label{eq:regrindler1}
\end{align}
where $-\infty<\eta<\infty$, $-\infty<\chi<\infty$, 
and $b$ is a positive constant. 
A~conformally flat form, 
obtained by the coordinate transformation $\chi = b \sinh y$, is 
\begin{align}
ds^2 = b^2 \cosh^2 \! y \, \bigl( - d\eta^2 + dy^2 \bigr) 
\ , 
\label{eq:regrindler2}
\end{align}
where
$-\infty<\eta<\infty$ and $-\infty<y<\infty$. 

The spacetime is curved, with the Ricci scalar 
\begin{align}
R = - \frac{2 b^2}{\bigl(b^2 + \chi^2 \bigr)^2} = - \frac{2}{b^2 \cosh^4 \! y}
\ . 
\label{eq:Ricci-scalar}
\end{align}
It is static, with the timelike Killing vector~$\partial_\eta$. 
The integral curves of $\partial_\eta$ are 
timelike worldlines of constant~$\chi$, and their proper acceleration is 
\begin{align}
a_\chi = \frac{|\chi|}{b^2 + \chi^2} 
\ . 
\label{eq:proper-acc}
\end{align}
It can be verified that the spacetime is geodesically complete. 

Comparison of \eqref{eq:unregrindler} and \eqref{eq:regrindler1}  
shows that the spacetime consists of 
asymptotically Rindler regions at $\chi\gg b$ and $\chi \ll -b$, 
joined by 
a negative curvature bridge whose effective length is of the order of~$b$.  
In the limit $b\to0$, \eqref{eq:regrindler1} reduces to~\eqref{eq:unregrindler}, 
which has exact Rindler wedges at $\chi>0$ and $\chi<0$ 
and a degeneracy at $\chi\to0_{\pm}$, where the Rindler horizons would be.  
We may think of $b$ as a regulator 
of the degeneracy of \eqref{eq:unregrindler} at the Rindler horizon.  
Note that the regularised spacetime does not have a Killing horizon since 
$\partial_\eta$ is everywhere timelike. 

The behaviour of $g_{\eta\eta}$ near $\chi=0$ in \eqref{eq:regrindler1} 
is reminiscent of the near-throat region of 
the $(2+1)$-dimensional and $(3+1)$-dimensional static wormhole spacetimes 
discussed in \cite{Solodukhin:2004rv,Solodukhin:2005qy,Damour:2007ap}, 
where a regulator analogous to our $b$ was introduced to remove a black hole Killing horizon. 
Interpreting \eqref{eq:regrindler1} as a wormhole spacetime 
would however be stretching the reminiscence 
because \eqref{eq:regrindler1} has no transverse dimensions  
whose size would attain a minimum at $\chi=0$.

\section{Quench\label{sec:quench}}

\subsection{Quench spacetime}

Our quench spacetime consists of the $\eta>0$ 
half of the regularised double Rindler spacetime \eqref{eq:regrindler1} 
joined to the $t<0$ half of Minkowski spacetime, 
given by 
\begin{align}
ds^2 = - dt^2 + dx^2
\ ,  
\end{align}
so that $x = \chi$ at $t=0=\eta$. 
The metric has a discontinuous time-time component at the quench 
but the other components are continuous. 

We wish to quantise a massless minimally coupled scalar 
field $\phi$ on the quench spacetime. The field equation is $\Box \phi=0$. 
We discuss the pre-quench region and the post-quench 
region first separately and then connect the two. 

\subsection{Quantum scalar field: pre-quench region}

In the pre-quench region, $t<0$, 
we employ a standard Fock quantisation adapted to the Killing vector~$\partial_t$.  
A~standard basis of mode functions that are 
positive frequency with respect to $\partial_t$ is 
\begin{align}
u_{\omega, \epsilon} = 
\frac{1}{\sqrt{4\pi\omega}} \exp[-i\omega(t - \epsilon x)]
\ , 
\label{eq:KGmodes-before}
\end{align}
where $\omega>0$ and $\epsilon \in \{1,-1\}$. The mode functions with 
$\epsilon=1$ are right-movers and the mode functions with 
$\epsilon=-1$ are left-movers. 
Adopting the conventions of~\cite{Birrell:1982ix},
the Klein-Gordon (indefinite) inner product on a constant $t$ hypersurface reads 
\begin{align}
(\phi, \psi) 
= - i \int_{-\infty}^{\infty} 
\left( \phi \partial_t \overline{\psi} - \overline{\psi} \partial_t \phi \right) 
dx 
\ ,  
\label{eq:KGip-before}
\end{align}
and the mode functions are normalised in this inner product to 
\begin{align}
(u_{\omega, \epsilon}, u_{\omega', \epsilon'}) 
= 
- (\overline{u}_{\omega, \epsilon}, \overline{u}_{\omega', \epsilon'}) 
= \delta_{\epsilon\epsilon'} \delta(\omega- \omega')
\ , 
\end{align}
with the mixed inner products vanishing. 

The quantised scalar field is expanded as 
\begin{align}
\phi = 
\sum_\epsilon \int_0^\infty 
\bigl( a_{\omega,\epsilon} u_{\omega,\epsilon} 
+ a^\dagger_{\omega,\epsilon} \overline{u}_{\omega,\epsilon} \bigr) 
\, d\omega
\ , 
\label{eq:phi-prequenchexpansion}
\end{align}
where $\bigl[a_{\omega,\epsilon}, a^\dagger_{\omega',\epsilon'}\bigr] 
= \delta_{\epsilon \epsilon'}\delta(\omega-\omega')$ 
and the other commutators vanish. The Fock space is built in the usual way on the 
Minkowski vacuum $|0_M\rangle$ that is normalised and satisfies 
$a_{\omega,\epsilon}|0_M\rangle =0$. 

\subsection{Quantum scalar field: post-quench region}

In the post-quench region, $\eta>0$, 
we employ a standard Fock quantisation adapted to the Killing vector~$\partial_\eta$.  
A~standard basis of mode functions that are 
positive frequency with respect to $\partial_\eta$ is 
\begin{align}
U_{\Omega, \epsilon} &= \frac{1}{\sqrt{4\pi\Omega}} 
\exp[-i\Omega(\eta - \epsilon y)]
\notag
\\
&= \frac{1}{\sqrt{4\pi\Omega}} 
\exp\bigl[-i\Omega\bigl(\eta - \epsilon \arsinh(\chi/b)\bigr)\bigr]
\ , 
\label{eq:post-quench-U}
\end{align}
where $\Omega>0$ and $\epsilon \in \{1,-1\}$. 
The mode functions with $\epsilon=1$ are again right-movers and the mode functions with  
$\epsilon=-1$ are left-movers. 
The Klein-Gordon inner product on a constant $\eta$ hypersurface reads 
\begin{align}
(\phi, \psi) 
= - i \int_{-\infty}^{\infty} 
\left( \phi \partial_\eta \overline{\psi} - \overline{\psi} \partial_\eta \phi \right) 
\frac{d\chi}{\sqrt{b^2+\chi^2}}
\ ,  
\label{eq:KGip-after}
\end{align}
and the mode functions are normalised in this inner product to 
\begin{align}
\bigl(U_{\Omega, \epsilon}, U_{\Omega', \epsilon'}\bigr) 
= - \bigl(\overline{U}_{\Omega, \epsilon}, \overline{U}_{\Omega', \epsilon'}\bigr) 
= \delta_{\epsilon\epsilon'} \delta(\Omega- \Omega')
\ , 
\end{align}
with the mixed inner products vanishing. 

The quantised scalar field is expanded as 
\begin{align}
\phi = 
\sum_\epsilon \int_0^\infty 
\bigl( A_{\Omega,\epsilon} U_{\Omega,\epsilon} 
+ A^\dagger_{\Omega,\epsilon} \overline{U}_{\Omega,\epsilon} \bigr) 
\, d\Omega
\ , 
\label{eq:phi-postquenchexpansion}
\end{align}
where $[A_{\Omega,\epsilon}, A^\dagger_{\Omega',\epsilon'}] 
= \delta_{\epsilon\epsilon'}\delta(\Omega-\Omega')$ 
and the other commutators vanish. 
A Fock space can be built in the usual way on the normalised state 
$|0_D\rangle$ that satisfies 
$A_{\Omega,\epsilon}|0_D\rangle =0$, and we may regard 
$|0_D\rangle$ as a regularised double-sided Rindler vacuum. 
In what follows we shall however not be interested in~$|0_D\rangle$ but instead in 
the post-quench state to which the pre-quench Minkowski vacuum evolves.

\section{Post-quench Wightman function\label{sec:wightman}}

We fix the state of the field to be the 
pre-quench Minkowski vacuum $|0_M\rangle$. 
To analyse the effects of the quench, 
we need to evaluate the Wightman function, 
\begin{align}
W(\sfx,\sfx') = \langle 0_M| \phi(\sfx) \phi(\sfx') |0_M\rangle
\ , 
\label{eq:Wightman-rawdef}
\end{align}
when both $\sfx$ and $\sfx'$ are in the post-quench region. 

We first match the pre-quench and post-quench mode functions at the quench, 
using the Bogoliubov transformation formalism in the conventions of~\cite{Birrell:1982ix}. 
Since the left-movers and right-movers decouple, 
we shall now drop the subscript~$\epsilon$, 
understand the transformation formulas to hold 
separately each value of~$\epsilon$, 
and add the left-mover and right-mover contributions to the Wightman function at the end. 

With this notation, we write the Bogoliubov transformation of the modes at $t=0=\eta$ as 
\begin{align}
U_\Omega = \int_0^\infty 
\left( \alpha_{\Omega\omega} u_\omega + \beta_{\Omega\omega} \overline{u}_\omega \right) 
d\omega 
\ , 
\label{eq:Bogo-def}
\end{align}
where $\alpha_{\Omega\omega}$ and $\beta_{\Omega\omega}$ 
are the Bogoliubov coefficients. 
The coefficients are given by \cite{Birrell:1982ix}
\begin{subequations}
\label{eq:bogocoeff-formulas}
\begin{align}
\alpha_{\Omega\omega} &= \bigl(U_\Omega,u_\omega \bigr)
\ , 
\\
\beta_{\Omega\omega} &= - \bigl(U_\Omega, \overline{u}_\omega \bigr)
\ , 
\end{align}
\end{subequations}
where the Klein-Gordon inner products are taken at the quench hypersurface $t=0=\eta$. 
Note that in these inner products the time derivative on 
$u_\omega$ is as in \eqref{eq:KGip-before} 
and the time derivative on 
$U_\Omega$ is as in~\eqref{eq:KGip-after}. 

The inner products in \eqref{eq:bogocoeff-formulas}
can be evaluated using 3.471.10 in~\cite{grad-ryzh}, with the result 
\begin{subequations}
\label{eq:bogo-coeffs}
\begin{align}
\alpha_{\Omega\omega} &= \frac{1}{\pi} \sqrt{\frac{\Omega}{\omega}} \, 
e^{\pi\Omega/2} K_{i\Omega}(b\omega)\ , 
\\
\beta_{\Omega\omega} &= - \frac{1}{\pi} \sqrt{\frac{\Omega}{\omega}} \, 
e^{-\pi\Omega/2} K_{i\Omega}(b\omega)\ , 
\end{align}
\end{subequations}
where $K$ is the modified Bessel function of the second kind. 
As a consistency check, it can be verified that 
the coefficients \eqref{eq:bogo-coeffs} satisfy 
the Bogoliubov identities~\cite{Birrell:1982ix}, using the 
Bessel function identity 
\begin{align}
\int_0^\infty \frac{dx}{x}K_{i\Omega}(x) K_{i\Omega'}(x) 
= 
\frac{\pi^2}{2\Omega \sinh(\pi\Omega)} \, \delta(\Omega-\Omega') 
\ . 
\label{eq:K-completeness}
\end{align}
\eqref{eq:K-completeness} can be justified informally by observing 
that $K_{i\Omega}(e^z)$ are the (improper) eigenfunctions of the 
essentially self-adjoint differential operator $-\partial^2_z + e^{2z}$, 
which implies orthogonality, and considering the small 
argument behaviour of $K_{i\Omega}(x)$~\cite{dlmf}, 
which determines the normalisation constant. 
A~rigorous discussion of \eqref{eq:K-completeness} is given in 
Section 4.15 of~\cite{titchmarsh-eigen1}. 

Next, we recall \cite{Birrell:1982ix} that 
\begin{align}
A_\Omega = \int_0^\infty 
\bigl(\alpha_{\Omega\omega} a_\omega -  \beta_{\Omega\omega} a^\dagger_\omega\bigr) 
\, d\omega
\ ,
\label{eq:A-ito-a}
\end{align}
where we have used the reality of the Bogoliubov coefficients. 
Proceeding for the moment informally, we substitute 
\eqref{eq:phi-postquenchexpansion} in~\eqref{eq:Wightman-rawdef}, 
use \eqref{eq:A-ito-a} and its Hermitian conjugate, interchange the integrals, 
and use the identity \eqref{eq:K-completeness}. 
Adding finally the left-mover and right-mover contributions, we arrive at 
\begin{align}
W(\sfx,\sfx') = W_0(\eta-y, \eta'-y') + W_0(\eta+y, \eta'+y')
\ , 
\label{eq:Wightman-decomp}
\end{align}
where 
\begin{align}
W_0(z, z') = 
\int_0^\infty \frac{d\Omega}{8 \pi \Omega \sinh(\pi\Omega)}
\left(e^{\pi\Omega/2} e^{-i\Omega z} + e^{-\pi\Omega/2} e^{i\Omega z}\right)
\left(e^{\pi\Omega/2} e^{i\Omega z'} + e^{-\pi\Omega/2} e^{-i\Omega z'}\right)
\ .
\label{eq:Wnought-div} 
\end{align}

The expression \eqref{eq:Wnought-div} for $W_0(z, z')$ is ill 
defined because of the small $\Omega$ behaviour of the integrand. 
This was to be expected because of the well known infrared ambiguity 
of the Wightman function of a massless scalar field in two dimensions~\cite{Decanini:2005eg}. 
To extract a meaningful expression for $W_0(z, z')$, 
we differentiate both sides of \eqref{eq:Wnought-div} with respect to $z$ 
and take the derivative on the right hand side to operate under the integral. 
The resulting integral has the distributional interpretation 
\begin{align}
\partial_z W_0(z, z') = 
- \frac{i}{4} \delta(z-z') 
- \frac{1}{8\pi} P \coth\!\left(\frac{z-z'}{2}\right)
- \frac{1}{8\pi} \tanh\!\left(\frac{z+z'}{2}\right)
\ ,
\label{eq:Wnought-partial-z} 
\end{align}
where $P$ stands for the Cauchy principal value and we have used 
3.981.1 and 3.981.8 in~\cite{grad-ryzh}. 
We now integrate \eqref{eq:Wnought-partial-z} with respect to~$z$, 
fixing the integration constant (which a priori could depend on~$z'$) by requiring 
$W_0(z, z') = \overline{W_0(z', z)}$, which a Wightman function must satisfy. 
We find 
\begin{align}
W_0(z, z') = 
- \frac{i}{8} \sgn(z-z') 
- \frac{1}{4\pi} \ln \! \left[\sinh\!\left(\frac{|z-z'|}{2}\right)\right]
- \frac{1}{4\pi} \ln \! \left[\cosh\!\left(\frac{z+z'}{2}\right)\right] 
\ ,    
\label{eq:Wnought-final} 
\end{align}
up to an additive purely numerical real-valued constant, 
which we have dropped from \eqref{eq:Wnought-final} 
as it will not affect what follows. 

To summarise, we have arrived at the post-quench 
Wightman function $W(\sfx, \sfx')$ given by 
\eqref{eq:Wightman-decomp} with \eqref{eq:Wnought-final}. 
The infrared divergence was removed by a procedure that 
can be interpreted as dropping an infinite additive constant. As a consistency check, 
we note that our $W(\sfx, \sfx')$ has the correct 
small separation asymptotic form~\cite{Decanini:2005eg}. 

We record here that the asymptotic late time form of 
$W(\sfx, \sfx')$ is 
\begin{align}
W_{\text{late}}(\sfx, \sfx') &= 
- \frac{i}{8} \sgn(\eta-\eta' + y - y')
- \frac{i}{8} \sgn(\eta-\eta' - y + y')
\notag
\\
&\hspace{3ex}
- \frac{1}{4\pi} \ln \! \left[\sinh\!\left(\frac{|\eta-\eta' + y - y'|}{2}\right)
\sinh\!\left(\frac{|\eta-\eta' - y + y'|}{2}\right)\right]
- \frac{\eta+\eta'}{2\pi}
\ . 
\label{eq:Wlate} 
\end{align}
We shall return to the implications of \eqref{eq:Wlate} in Section~\ref{sec:discussion}.

\section{Post-quench stress-energy\label{sec:stress-energy}}

We now evaluate the post-quench renormalised stress-energy tensor. 

We use Hadamard renormalisation, adapting the 
Feynman Green's function formalism of \cite{Decanini:2005eg} to the Wightman function. This gives 
\begin{align}
T_{ab}(\sfx) = 
\lim_{\sfx' \to \sfx}
\left(g_b{}^{b'}\partial_a\partial_{b'} - \tfrac12 g_{ab} g^{cd'}\partial_c \partial_{d'}\right)
\left(W(\sfx,\sfx') - W_{\text{sing}}(\sfx,\sfx')\right)
+ \frac{1}{48\pi} R(\sfx) g_{ab}
\ , 
\end{align}
where the purely geometric subtraction term is 
\begin{align}
W_{\text{sing}}(\sfx,\sfx')
= - \frac{1}{4\pi}\ln|\sigma(\sfx,\sfx')| 
- \frac{i}{8} \sgn(\eta-\eta' + y-y') 
- \frac{i}{8} \sgn(\eta-\eta' - y+y') 
\ , 
\end{align}
and $\sigma(\sfx,\sfx')$ is half of the geodesic 
distance squared between~$\sfx$ and~$\sfx'$, 
with the convention that $\sigma(\sfx,\sfx')>0$ when the geodesic is spacelike and 
$\sigma(\sfx,\sfx')<0$ when the geodesic is timelike.
For a metric of the form $ds^2 = F(y) \bigl(- d\eta^2 + dy^2\bigr)$, 
a small separation expansion gives 
\begin{align}
\sigma(\sfx,\sfx') &= \tfrac12 \! \left( (y-y')^2 - (\eta-\eta')^2 \right)
F(\tilde y)
\notag
\\
& \hspace{1ex}
\times 
\left[1 + \frac{F''(\tilde y)}{24 F(\tilde y)} (y-y')^2 
- \frac{1}{48}\left(\frac{F'(\tilde y)}{F(\tilde y)}\right)^2 
\left( (y-y')^2 - (\eta-\eta')^2 \right) 
+ \text{(cubic)}\right]
\ , 
\label{eq:twosigma-short}
\end{align}
where $\tilde y := (y + y')/2$. 
Using \eqref{eq:twosigma-short} with 
$F(y) = b^2 \cosh^2\!y$, the Wightman function given 
by \eqref{eq:Wightman-decomp} and~\eqref{eq:Wnought-final}, 
and the Ricci scalar~\eqref{eq:Ricci-scalar}, we find 
\begin{subequations}
\label{eq:Tfinal-y}
\begin{align}
T_{\eta\eta} &= 
\frac{1}{8\pi\cosh^2\!y}
- 
\frac{1}{16\pi} \!
\left(\frac{1}{\cosh^2(\eta-y)} + \frac{1}{\cosh^2(\eta+y)}\right)
\ , 
\label{eq:Tetaeta}
\\
T_{yy} &= 
\frac{1}{24\pi\cosh^2\!y}
- 
\frac{1}{16\pi} \!
\left(\frac{1}{\cosh^2(\eta-y)} + \frac{1}{\cosh^2(\eta+y)}\right)
\ , 
\\
T_{\eta y} &= 
\frac{1}{16\pi} \!
\left(\frac{1}{\cosh^2(\eta-y)} - \frac{1}{\cosh^2(\eta+y)}\right)
\ . 
\end{align}
\end{subequations}
$T_{ab}$ is hence well defined and finite everywhere in the post-quench region. 
As a consistency check, 
it can be verified that $T_{ab}$ is conserved, $\nabla_a T^{ab}=0$, 
and it has the correct trace anomaly, 
$T^a{}_a = R/(24\pi)$~\cite{Decanini:2005eg}. 

From the expressions in \eqref{eq:Tfinal-y} 
we may make the following three observations. 

First, at the quench, $\eta\to0_+$, we have 
$T_{yy} \to - 1/(12 \pi \cosh^2 \! y)$, while the other components vanish. 
The quench creates initially a negative pressure but no energy density. 

Second, in the evolution after the quench, 
$T_{\eta\eta}$ and $T_{yy}$ 
each consist of a positive static contribution, peaked around $y=0$, 
and negative pulses travelling to the left 
and right at the speed of light, 
peaked around $y=\pm\eta$. 
In the late time limit at fixed~$y$, 
the pulses have passed, and we have 
$T_{\eta\eta} \to 1/(8 \pi \cosh^2 \! y)$, 
$T_{yy} \to 1/(24 \pi \cosh^2 \! y)$ and $T_{\eta y} \to 0$. 
For fixed~$y$, the late time energy density and pressure 
are hence static and positive. 

Third, in view of the analogue system of~\cite{Rodriguez-Laguna:2016kri,Kosior:2018vgx}, 
an energetic quantity of interest is the `total energy' at constant~$\eta$, 
defined as the integral of the energy density 
$-T^\eta{}_\eta$ over the spatial volume, 
\begin{align}
E_\eta &:= - \int_{-\infty}^\infty T^\eta{}_\eta \, b \cosh y \, dy 
\notag
\\
&= \frac{\tanh^2(\eta/2)}{16\pi b} 
\ . 
\label{eq:E-sub-eta}
\end{align}
$E_\eta$ is finite for all~$\eta$, and it increases monotonically from 
$0$ to $1/(16\pi b)$ as $\eta$ increases from $0$ to infinity. 
The initial negative pressure hence evolves at late times 
into a finite and static positive total energy. 

If we view the parameter $b$ as a regulator that is small 
compared with length scales of interest, 
it is useful to express $T_{ab}$ in the 
coordinates $(\eta,\chi)$ of \eqref{eq:regrindler1}, with the result 
\begin{subequations}
\label{eq:Tfinal-chi}
\begin{align}
T_{\eta\eta} &= 
\frac{b^2}{8\pi}
\left(\frac{1}{\chi^2+b^2} 
- \frac{\chi^2 \cosh(2\eta) + b^2 \cosh^2 \! \eta}{\bigl(\chi^2+b^2 \cosh^2 \! \eta \bigr)^2}\right)
\ , 
\label{eq:Tetaeta-chi}
\\
T_{\chi\chi} &= 
\frac{b^2}{8\pi\bigl(\chi^2+b^2\bigr)}
\left(\frac{1}{3 \bigl(\chi^2+b^2\bigr)} 
- \frac{\chi^2 \cosh(2\eta) + b^2 \cosh^2 \! \eta}{\bigl(\chi^2+b^2 \cosh^2 \! \eta \bigr)^2}\right)
\ , 
\\
T_{\eta \chi} &= 
\frac{b^2}{8\pi}
\frac{\chi \sinh(2\eta)}
{\bigl(\chi^2+b^2 \cosh^2 \! \eta \bigr)^2}
\ . 
\end{align}
\end{subequations}
In these coordinates, the pointwise limit of $T_{ab}$ as $b\to0$ 
vanishes for $\chi\ne0$ but diverges for $\chi=0$. For $b$ small but finite, 
$T_{ab}$ is large only within the narrow region $|\chi| \lesssim b \cosh\eta$, 
and in particular it is this narrow region that contributes to the total energy 
$E_\eta$ \eqref{eq:E-sub-eta} the piece that diverges as $b\to0$. 

In summary, the regularised quench produces a well-defined 
stress-energy tensor everywhere to the future of the quench. 
When the regulator is small, the stress-energy tensor is 
small everywhere except in a narrow wedge about $\chi=0$.

\section{Post-quench thermality\label{sec:thermality}}

To examine thermality in the 
post-quench region, we probe the field with a pointlike 
Unruh-DeWitt detector \cite{Unruh:1976db,DeWitt:1979}, 
specifically with a variant that is coupled linearly to the field's proper 
time derivative rather than the field itself, 
since this makes the detector less sensitive to the 
infrared ambiguity in the Wightman function 
(for selected references 
see \cite{Grove:1986fy,Raine:1991kc,Raval:1995mb,Davies:2002bg,Wang:2013lex,Juarez-Aubry:2014jba}). 
We follow the notation of~\cite{Juarez-Aubry:2014jba}, to which we refer for the details. 

We take the detector to follow a worldline of constant~$\chi$, 
that is, an orbit of the Killing vector~$\partial_\eta$. 
Let $\tau$ be the proper time on this worldline, with the additive constant 
chosen so that $\tau=0$ at the quench. The detector's response is 
determined by the pull-back of the Wightman 
function on this worldline, given by 
\begin{align}
W_\chi (\tau,\tau') &= 
- \frac{i}{4} \sgn(\tau-\tau')
- \frac{1}{2\pi}
\ln\!\left[\sinh \! \left(\frac{|\tau-\tau'|}{2\sqrt{\chi^2+b^2}}\right)\right]
\notag
\\
& \hspace{3ex}
- \frac{1}{4\pi}
\ln \! \left[\cosh \! \left(\frac{\tau+\tau'}{\sqrt{\chi^2+b^2}}\right) + 1 + \frac{2\chi^2}{b^2}\right]
\ , 
\label{eq:W-pullback}
\end{align}
where we have dropped an additive numerical constant. 
Comparing \eqref{eq:W-pullback} to Section 3.3 in \cite{Juarez-Aubry:2014jba} 
shows that if the last term in \eqref{eq:W-pullback} can be neglected, 
and the detector operates so long that switch-on and switch-off effects 
are negligible, the transition rate, 
evaluated to first order in perturbation theory and dropping an overall multiplicative constant, 
takes the Planckian form  
\begin{align}
\dot{\mathcal F}(E) = \frac{E}{e^{E/T_\chi}-1}
\ , 
\label{eq:Fdot-Planckian}
\end{align}
where $E$ is the detector's energy gap and
\begin{align}
T_\chi 
= \frac{1}{2\pi \sqrt{b^2+\chi^2}}
\ . 
\label{eq:temperature}
\end{align}
When \eqref{eq:Fdot-Planckian} holds, 
$\dot{\mathcal F}$ is hence thermal in temperature~$T_\chi$, 
in the sense of the detailed balance condition,  
\begin{align}
\dot{\mathcal F}(-E) = e^{E/T_\chi}\dot{\mathcal F}(E)
\ . 
\end{align}

We note that when $|\chi|\to\infty$ with fixed~$b$, 
$T_\chi$ is asymptotically equal to~$a_\chi/(2\pi)$, 
where $a_\chi$ is the trajectory's proper acceleration~\eqref{eq:proper-acc}; 
conversely, for fixed $\chi\ne0$, 
taking the regulator $b$ to zero makes both $T_\chi$ 
and $a_\chi/(2\pi)$ tend to $1/(2\pi|\chi|)$, 
which is the Unruh temperature on the Rindler trajectory of constant $\chi\ne0$ 
in the unregularised Rindler metric~\eqref{eq:unregrindler}. 
When \eqref{eq:Fdot-Planckian} holds, 
the regularised quench hence makes the detector 
respond identically to the Unruh effect, at scales that are 
large compared with the regulator~$b$. 

Now, when does \eqref{eq:Fdot-Planckian} hold? That is, 
when does the last term in \eqref{eq:W-pullback} make a negligible contribution to 
$\partial_\tau \partial_{\tau'} W_\chi (\tau,\tau')$? 
For any fixed $\chi$ and~$b$, it is clear from \eqref{eq:W-pullback} that 
one regime where this happens 
is the late time limit. However, if we view $b$ as a regulator 
that is small compared with length scales of interest, the situation to consider is to fix 
$\chi\ne0$ and take $b\ll |\chi|$. 
The late time limit in which \eqref{eq:Fdot-Planckian} 
holds is then at proper times much larger than $|\chi|\ln(2|\chi|/b)$. 
But \eqref{eq:W-pullback} shows that 
\eqref{eq:Fdot-Planckian} then holds 
also at early post-quench proper times, 
much smaller than~$|\chi|\ln(2|\chi|/b)$. 
This might have been expected from the stress-energy 
analysis of Section~\ref{sec:stress-energy}, 
since $|\chi|\ln(2|\chi|/b)$ 
is the proper time at which the detector crosses 
a travelling peak in~$T_{\eta\eta}$. 

We conclude that when $\chi\ne0$ and $b\ll |\chi|$, the detector's transition 
rate is approximately Planckian 
at approximately the usual Unruh temperature $1/(2\pi|\chi|)$ 
at proper times much larger and much smaller than $|\chi|\ln(2|\chi|/b)$. 
The sense of the approximations can be made precise using 
\eqref{eq:W-pullback} and the transition rate formalism of~\cite{Juarez-Aubry:2014jba}. 
Inclusion of finite time switch-on and switch-off effects would be 
analytically more involved~(cf.~\cite{Fewster:2016ewy}), 
but straightforward to implement numerically.

\section{Summary and discussion\label{sec:discussion}}

We have provided new support for the proposal of \cite{Rodriguez-Laguna:2016kri,Kosior:2018vgx} 
to simulate the Unruh effect experimentally with ultracold fermionic atoms in an optical lattice. 
We first identified three qualitative differences between 
the optical lattice system and the usual Unruh effect setup, 
in their energetic and causal properties, and in the fact 
that the lattice provides a horizon regulator that has no 
counterpart in the usual Unruh effect. 
These differences could cast doubt on the ability of the 
lattice to simulate the Unruh effect. 
We then presented a simplified continuum field theory model 
that shares the potentially troublesome features of the 
optical lattice, and showed that in this model the energetic 
and causal properties can be brought under analytic control, 
and the Unruh effect prevails. 
While our simplifications included going from effective spacetime 
dimension $(2+1)$ to effective spacetime dimension $(1+1)$, 
and replacing a discrete fermion field by a continuum scalar field,
our analytic results are compatible with the analytic and numerical conclusions 
obtained in~\cite{Rodriguez-Laguna:2016kri,Kosior:2018vgx}. 

In summary, our results strengthen the motivation to realise 
the experimental proposal of~\cite{Rodriguez-Laguna:2016kri,Kosior:2018vgx}. 

A key technical property that made our analysis feasible 
was that the Wightman function could be written down in closed form, 
and we used this Wightman function to evaluate the 
stress-energy tensor and to establish the thermal response 
of a static Unruh-DeWitt detector. 
Given the Wightman function, it would be possible to 
study also the spatial entanglement in the field, 
harvesting the entanglement by a pair of Unruh-DeWitt 
detectors~\cite{Salton:2014jaa,valentini1991,Reznik:2002fz,Reznik:2003mnx,Pozas-Kerstjens:2015gta}, 
and to compare with the entanglement that is present 
in Minkowski vacuum for Rindler observers in opposing Rindler 
wedges~\cite{Fulling:1972md,Davies:1974th,Unruh:1976db,Birrell:1982ix,Crispino:2007eb,Fulling:2014wzx}. 
Because of the late time growth in the Wightman function, shown in~\eqref{eq:Wlate}, 
a pair of Unruh-DeWitt detectors coupled linearly to the field would be problematic. 
A~pair of Unruh-DeWitt detectors coupled linearly to the proper time derivative of the field, 
used in Section~\ref{sec:thermality}, 
would avoid this problem, but the short distance properties 
of the twice differentiated Wightman function then require the detectors 
to be smeared in time and space~\cite{Pozas-Kerstjens:2015gta}, 
increasing the parameter space of the harvesting protocol, 
and suggesting the need for a numerical approach. 
We leave this question to future work.

\section*{Acknowledgments}

This work originated at a June 2018 Unruh effect workshop 
organised at the University of 
Nottingham by Silke Weinfurtner, supported by FQXi 
(Mini-Grant FQXi-MGB-1742 ``Detecting Unruh Radiation''). 
I~thank the workshop participants, particularly Alessio Celi, 
for stimulating discussions. 
I~also thank 
Uwe Fischer, 
Sergey Solodukhin 
and 
Husni Wan Mokhtar 
for bringing related work to my attention
and an anonymous referee for helpful presentational suggestions. 
This work was supported in part by 
Science and Technology Facilities Council 
(Theory Consolidated Grant ST/P000703/1). 

\vspace{2ex}

\emph{Note added in proof:} 
A metric obtained from \eqref{eq:regrindler1} by continuing $b^2$ to
negative values has been considered in \cite{Caianiello:1989wm,Caianiello:1989pu} 
as a consequence of an
upper bound on proper acceleration. 
I~thank Maurizio Gasperini for
bringing this work to my attention.


\begin{thebibliography}{99}

\bibitem{Fulling:1972md} 
S.~A.~Fulling,
``Nonuniqueness of canonical field quantization in Riemannian space-time,''
Phys.\ Rev.\ D {\bf 7}, 2850 (1973).

\bibitem{Davies:1974th} 
P.~C.~W.~Davies,
``Scalar particle production in Schwarzschild and Rindler metrics,''
J.\ Phys.\ A {\bf 8}, 609 (1975).

\bibitem{Unruh:1976db} 
W.~G.~Unruh,
``Notes on black hole evaporation,''
Phys.\ Rev.\ D {\bf 14}, 870 (1976).

\bibitem{Birrell:1982ix}
N.~D.~Birrell
and
P.~C.~W.~Davies, 
\textit{Quantum Fields in Curved Space} 
(Cambridge University Press, Cambridge, 1982).

\bibitem{Crispino:2007eb}
L.~C.~B.~Crispino, A.~Higuchi and G.~E.~A.~Matsas,
``The Unruh effect and its applications,''
Rev.\ Mod.\ Phys.\  {\bf 80}, 787 (2008)
[arXiv:0710.5373 [gr-qc]].

\bibitem{Fulling:2014wzx} 
S.~Fulling and G.~Matsas,
``Unruh effect,''
Scholarpedia {\bf 9}, no. 10, 31789 (2014).

\bibitem{MartinMartinez:2010sg} 
E.~Mart\'in-Mart\'inez, I.~Fuentes and R.~B.~Mann,
``Using Berry's phase to detect the Unruh effect at lower accelerations,''
Phys.\ Rev.\ Lett.\  {\bf 107}, 131301 (2011)
[arXiv:1012.2208 [quant-ph]].

\bibitem{Letaw:1980yv} 
J.~R.~Letaw,
``Vacuum Excitation of Noninertial Detectors on Stationary World Lines,''
Phys.\ Rev.\ D {\bf 23}, 1709 (1981).

\bibitem{Letaw:1979wy} 
J.~R.~Letaw and J.~D.~Pfautsch,
``The Quantized Scalar Field in Rotating Coordinates,''
Phys.\ Rev.\ D {\bf 22}, 1345 (1980).

\bibitem{Doukas:2010wt} 
J.~Doukas and B.~Carson,
``Entanglement of two qubits in a relativistic orbit,''
Phys.\ Rev.\ A {\bf 81}, 062320 (2010)
[arXiv:1003.2201 [quant-ph]].

\bibitem{Jin:2014coa} 
Y.~Jin, J.~Hu and H.~Yu,
``Spontaneous excitation of a circularly accelerated atom 
coupled to electromagnetic vacuum fluctuations,''
Annals Phys.\  {\bf 344}, 97 (2014).

\bibitem{Jin:2014spa} 
Y.~Jin, J.~Hu and H.~Yu,
``Dynamical behavior and geometric phase for a circularly accelerated two-level atom,''
Phys.\ Rev.\ A {\bf 89}, 
064101 (2014)
[arXiv:1406.5576 [gr-qc]].

\bibitem{Bell:1982qr} 
J.~S.~Bell and J.~M.~Leinaas,
``Electrons as accelerated thermometers,''
Nucl.\ Phys.\ B {\bf 212}, 131 (1983).

\bibitem{Bell:1986ir} 
J.~S.~Bell and J.~M.~Leinaas,
``The Unruh effect and quantum fluctuations of electrons in storage rings,''
Nucl.\ Phys.\ B {\bf 284}, 488 (1987).

\bibitem{Leinaas:1998tu} 
J.~M.~Leinaas,
``Accelerated electrons and the Unruh effect,''
in 
{\it Quantum aspects of beam physics. 
Proceedings, Advanced ICFA Beam Dynamics Workshop, Monterey, USA, January 4-9, 1998}, 
edited by P.~Chen 
(World Scientific, Singapore, 1999).

\bibitem{Unruh:1998gq} 
W.~G.~Unruh,
``Acceleration radiation for orbiting electrons,''
Phys.\ Rept.\  {\bf 307}, 163 (1998)
[arXiv:hep-th/9804158].

\bibitem{Sokolov:1963zn} 
A.~A.~Sokolov and I.~M.~Ternov,
``On polarization and spin effects in the theory of synchrotron radiation,''
Sov.\ Phys.\ Dokl.\  {\bf 8}, 1203 (1964)
[Dokl.\ Akad.\ Nauk Ser.\ Fiz.\  {\bf 153}, 1052 (1964)]. 


\bibitem{Jackson:1975qi} 
J.~D.~Jackson,
``On Understanding Spin-Flip Synchrotron Radiation and the 
Transverse Polarization of Electrons in Storage Rings,''
Rev.\ Mod.\ Phys.\  {\bf 48}, 417 (1976).

\bibitem{Johnson:1982rb} 
J.~R.~Johnson, R.~Prepost, D.~E.~Wiser, J.~J.~Murray, R.~Schwitters and C.~K.~Sinclair,
``Beam Polarization Measurements at the {SPEAR} Storage Ring,''
Nucl.\ Instrum.\ Meth.\  {\bf 204}, 261 (1983).

\bibitem{Chen:1998kp} 
P.~Chen and T.~Tajima,
``Testing Unruh radiation with ultraintense lasers,''
Phys.\ Rev.\ Lett.\  {\bf 83}, 256 (1999).

\bibitem{Mourou:2006zz} 
G.~A.~Mourou, T.~Tajima and S.~V.~Bulanov,
``Optics in the relativistic regime,''
Rev.\ Mod.\ Phys.\  {\bf 78}, 309 (2006).

\bibitem{Brodin:2007vf} 
G.~Brodin, M.~Marklund, R.~Bingham, J.~Collier and R.~G.~Evans,
``Laboratory soft x-ray emission due to the Hawking-Unruh effect?,''
Class.\ Quant.\ Grav.\  {\bf 25}, 145005 (2008)
[arXiv:0712.2985 [hep-ph]].

\bibitem{gregori-scirep} 
B. J. B. Crowley {\it et al\/}, 
``Testing quantum mechanics in non-Minkowski space-time with high power 
lasers and 4th generation light sources,''
Sci.\ Rep.\ {\bf 2}, 491 (2012). 

\bibitem{Hawking:1974sw} 
S.~W.~Hawking,
``Particle Creation by Black Holes,''
Commun.\ Math.\ Phys.\  {\bf 43}, 199 (1975)
[Erratum-ibid.\  {\bf 46}, 206 (1976)].

\bibitem{Parker:1969au} 
L.~Parker,
``Quantized fields and particle creation in expanding universes. 1.,''
Phys.\ Rev.\  {\bf 183}, 1057 (1969).

\bibitem{mukha-wini-book}
V.~Mukhanov and S.~Winitzki, 
\textit{Introduction to Quantum Effects in Gravity} 
(Cambridge University Press, Cambridge, 2007).

\bibitem{Barcelo:2005fc} 
C.~Barcelo, S.~Liberati and M.~Visser,
``Analogue gravity,''
Living Rev.\ Rel.\  {\bf 8}, 12 (2005)
[Living Rev.\ Rel.\  {\bf 14}, 3 (2011)]
[arXiv:gr-qc/0505065].

\bibitem{Weinfurtner:2010nu} 
S.~Weinfurtner, E.~W.~Tedford, M.~C.~J.~Penrice, W.~G.~Unruh and G.~A.~Lawrence,
``Measurement of stimulated Hawking emission in an analogue system,''
Phys.\ Rev.\ Lett.\  {\bf 106}, 021302 (2011)
[arXiv:1008.1911 [gr-qc]].

\bibitem{Torres:2016iee} 
T.~Torres, S.~Patrick, A.~Coutant, M.~Richartz, E.~W.~Tedford and S.~Weinfurtner,
``Rotational superradiant scattering in a vortex flow,''
Nature Phys.\  {\bf 13}, 833 (2017)
[arXiv:1612.06180 [gr-qc]].

\bibitem{Eckel:2017uqx} 
S.~Eckel, A.~Kumar, T.~Jacobson, I.~B.~Spielman and G.~K.~Campbell,
``A rapidly expanding Bose-Einstein condensate: an expanding universe in the lab,''
Phys.\ Rev.\ X {\bf 8}, 021021 (2018)
[arXiv:1710.05800 [cond-mat.quant-gas]].

\bibitem{Belgiorno:2010wn} 
F.~Belgiorno {\it et al.},
``Hawking radiation from ultrashort laser pulse filaments,''
Phys.\ Rev.\ Lett.\  {\bf 105}, 203901 (2010)
[arXiv:1009.4634 [gr-qc]].

\bibitem{Steinhauer:2015saa} 
J.~Steinhauer,
``Observation of quantum Hawking radiation and its entanglement in an analogue black hole,''
Nature Phys.\  {\bf 12}, 959 (2016)
[arXiv:1510.00621 [gr-qc]].

\bibitem{Retzker-circular} 
A.~Retzker, 
J.~I.~Cirac, 
M.~B.~Plenio
and 
B.~Reznik, 
``Methods for detecting acceleration radiation in a Bose-Einstein condensate,''
Phys.\ Rev.\ Lett.\ {\bf 101}, 110402 (2008)
[arXiv:0709.2425 [quant-ph]].

\bibitem{Rodriguez-Laguna:2016kri} 
J.~Rodr\'iguez-Laguna, L.~Tarruell, M.~Lewenstein and A.~Celi,
``Synthetic Unruh effect in cold atoms,''
Phys.\ Rev.\ A {\bf 95}, 013627 (2017)
[arXiv:1606.09505 [cond-mat.quant-gas]].

\bibitem{Kosior:2018vgx} 
A.~Kosior, M.~Lewenstein and A.~Celi,
``Unruh effect for interacting particles with ultracold atoms,''
arXiv:1804.11323 [cond-mat.quant-gas].

\bibitem{Cozzella:2017ckb} 
G.~Cozzella, A.~G.~S.~Landulfo, G.~E.~A.~Matsas and D.~A.~T.~Vanzella,
``Proposal for Observing the Unruh Effect using Classical Electrodynamics,''
Phys.\ Rev.\ Lett.\  {\bf 118}, 161102 (2017)
[arXiv:1701.03446 [gr-qc]].

\bibitem{Leonhardt:2017lwm} 
U.~Leonhardt, I.~Griniasty, S.~Wildeman, E.~Fort and M.~Fink,
``Classical analog of the Unruh effect,''
Phys.\ Rev.\ A {\bf 98}, 022118 (2018)
[arXiv:1709.02200 [gr-qc]].

\bibitem{Fedichev:2003id} 
P.~O.~Fedichev and U.~R.~Fischer,
``Gibbons-Hawking effect in the sonic de Sitter space-time of an 
expanding Bose-Einstein-condensed gas,''
Phys.\ Rev.\ Lett.\  {\bf 91}, 240407 (2003)
[arXiv:cond-mat/0304342].

\bibitem{Fedichev:2003dj} 
P.~O.~Fedichev and U.~R.~Fischer,
``Observer dependence for the phonon content of the sound field living 
on the effective curved space-time background of a Bose-Einstein condensate,''
Phys.\ Rev.\ D {\bf 69}, 064021 (2004)
[arXiv:cond-mat/0307200].

\bibitem{Buchholz:2014jta} 
D.~Buchholz and R.~Verch,
``Macroscopic aspects of the Unruh effect,''
Class.\ Quant.\ Grav.\  {\bf 32}, 245004 (2015)
[arXiv:1412.5892 [gr-qc]].

\bibitem{Buchholz:2015fqa}
D.~Buchholz and R.~Verch,
``Unruh versus Tolman: On the heat of acceleration,''
Gen.\ Rel.\ Grav.\  {\bf 48}, 32 (2016)
[arXiv:1505.01686 [gr-qc]].

\bibitem{Salton:2014jaa} 
G.~Salton, R.~B.~Mann and N.~C.~Menicucci,
``Acceleration-assisted entanglement harvesting and rangefinding,''
New J.\ Phys.\  {\bf 17}, 035001 (2015)
[arXiv:1408.1395 [quant-ph]].

\bibitem{valentini1991} 
A.~Valentini, 
``Non-local correlations in quantum electrodynamics,''
Phys.\ Lett.\ A {\bf 153}, 321 (1991).

\bibitem{Reznik:2002fz} 
B.~Reznik,
``Entanglement from the vacuum,''
Found.\ Phys.\  {\bf 33}, 167 (2003)
[arXiv:quant-ph/0212044].

\bibitem{Reznik:2003mnx} 
B.~Reznik, A.~Retzker and J.~Silman,
``Violating Bell's inequalities in the vacuum,''
Phys.\ Rev.\ A {\bf 71}, 042104 (2005)
[arXiv:quant-ph/0310058].

\bibitem{Pozas-Kerstjens:2015gta} 
A.~Pozas-Kerstjens and E.~Mart\'in-Mart\'inez,
``Harvesting correlations from the quantum vacuum,''
Phys.\ Rev.\ D {\bf 92}, 064042 (2015)
[arXiv:1506.03081 [quant-ph]].

\bibitem{Solodukhin:2004rv} 
S.~N.~Solodukhin,
``Can black hole relax unitarily?,''
in 
{\it Mathematical, Theoretical and Phenomenological Challenges Beyond the Standard Model: 
Perspectives of the Balkan Collaborations}, 
edited by 
G.~Djordjevi\'c, 
L.~Ne\v{s}i\'c, 
and J.~Wess (World Scientific, Singapore, 2005)
[arXiv:hep-th/0406130].

\bibitem{Solodukhin:2005qy} 
S.~N.~Solodukhin,
``Restoring unitarity in BTZ black hole,''
Phys.\ Rev.\ D {\bf 71}, 064006 (2005)
[arXiv:hep-th/0501053].

\bibitem{Damour:2007ap} 
T.~Damour and S.~N.~Solodukhin,
``Wormholes as black hole foils,''
Phys.\ Rev.\ D {\bf 76}, 024016 (2007)
[arXiv:0704.2667 [gr-qc]].

\bibitem{grad-ryzh}
I.~S.~Gradshteyn 
and 
I.~M.~Ryzhik,
\textit{Table of Integrals, Series, and Products\/},
7th edition 
(Academic Press, New York, 2007).

\bibitem{dlmf} 
\textit{NIST Digital Library of Mathematical Functions.} 
\texttt{http://dlmf.nist.gov/}, Release 1.0.19 of 2018-06-22. 
F.~W.~J. Olver, A.~B. Olde Daalhuis, D.~W. Lozier, 
B.~I. Schneider, R.~F. Boisvert, C.~W. Clark, 
B.~R. Miller, and B.~V. Saunders, eds.

\bibitem{titchmarsh-eigen1} 
E.~C. Titchmarsh, 
\textit{Eigenfunction Expansions\/}, Part 1, 2nd edition 
(Oxford University Press, Oxford, 1962).  

\bibitem{Decanini:2005eg} 
Y.~D\'ecanini and A.~Folacci,
``Hadamard renormalization of the stress-energy tensor for a quantized 
scalar field in a general spacetime of arbitrary dimension,''
Phys.\ Rev.\ D {\bf 78}, 044025 (2008)
[arXiv:gr-qc/0512118].

\bibitem{DeWitt:1979}
B.~S.~DeWitt,
``Quantum gravity: the new synthesis'', 
in 
{\it General Relativity: an Einstein centenary survey}, 
edited by S.~W.~Hawking and W.~Israel 
(Cambridge University Press, Cambridge, 1979).

\bibitem{Grove:1986fy} 
P.~G.~Grove,
``On the Detection of Particle and Energy Fluxes in Two-dimensions,''
Class.\ Quant.\ Grav.\  {\bf 3}, 793 (1986).

\bibitem{Raine:1991kc} 
D.~J.~Raine, D.~W.~Sciama and P.~G.~Grove,
``Does an accelerated oscillator radiate?''
Proc.\ Roy.\ Soc.\ A {\bf 435}, 205
(1991).

\bibitem{Raval:1995mb} 
A.~Raval, B.~L.~Hu and J.~Anglin,
``Stochastic theory of accelerated detectors in a quantum field,''
Phys.\ Rev.\ D {\bf 53}, 7003 (1996)
[arXiv:gr-qc/9510002].

\bibitem{Davies:2002bg}
P.~C.~W.~Davies and A.~C.~Ottewill,
``Detection of negative energy: 4-dimensional examples,''
Phys.\ Rev.\ D {\bf 65}, 104014 (2002) 
[arXiv:gr-qc/0203003].

\bibitem{Wang:2013lex} 
Q.~Wang and W.~G.~Unruh,
``Motion of a mirror under infinitely fluctuating quantum vacuum stress,''
Phys.\ Rev.\ D {\bf 89}, 085009 (2014)
[arXiv:1312.4591 [gr-qc]].

\bibitem{Juarez-Aubry:2014jba} 
B.~A.~Ju\'arez-Aubry and J.~Louko,
``Onset and decay of the 1 + 1 Hawking-Unruh effect: 
what the derivative-coupling detector saw,''
Class.\ Quant.\ Grav.\  {\bf 31}, 245007 (2014)
[arXiv:1406.2574 [gr-qc]].

\bibitem{Fewster:2016ewy} 
C.~J.~Fewster, B.~A.~Ju\'arez-Aubry and J.~Louko,
``Waiting for Unruh,''
Class.\ Quant.\ Grav.\  {\bf 33}, 165003 (2016)
[arXiv:1605.01316 [gr-qc]].

\bibitem{Caianiello:1989wm} 
E.~R.~Caianiello, A.~Feoli, M.~Gasperini and G.~Scarpetta,
``Quantum Corrections to the Space-time Metric From Geometric Phase Space Quantization,''
Int.\ J.\ Theor.\ Phys.\  {\bf 29}, 131 (1990).

\bibitem{Caianiello:1989pu} 
E.~R.~Caianiello, M.~Gasperini and G.~Scarpetta,
``Phenomenological Consequences of a Geometric Model With Limited Proper Acceleration,''
Nuovo Cim.\ B {\bf 105}, 259 (1990).


\end{thebibliography}
\end{document}